\pdfoutput=1 
\documentclass[]{JINST}

\title{Search for New Physics in SHiP and at future colliders}

\author{E. Graverini$^a$\thanks{Corresponding author.},
N. Serra$^a$
and B. Storaci$^a$, on behalf of the SHiP collaboration\\
\llap{$^a$} Physik-Institut, University of Zurich,

Winterthurerstrasse 190, Zurich, Switzerland\\
E-mail: \email{elena.graverini@cern.ch}\\

\textbf{Proceeding of the 2$^{nd}$ International Summer School on Intelligent Signal Processing for Frontier Research and Industry}}

\abstract{SHiP is a newly proposed fixed-target experiment at the CERN SPS with the aim of searching for hidden particles that interact very weakly with Standard Model particles. The work presented in this document investigates SHiP's physics reach in the parameter space of the Neutrino Minimal Standard Model ($\nu$MSM), a theory that could solve most problems unexplained by the Standard Model by incorporating sterile neutrinos. A model introducing an extra $U(1)$ symmetry in the hidden sector, providing a natural candidate for dark matter, is also explored. This work shows that the SHiP experiment can improve the sensitivity to Heavy Neutral Leptons below 2~GeV by several orders of magnitude, scanning a large part of the parameter space below the $B$ meson mass. The remainder of the $\nu$MSM parameter space, dominated by right-handed neutrinos with masses above 2~GeV, can be explored at a future $e^+e^-$ collider. Similarly, SHiP can greatly improve present constraints on $U(1)$ dark photons.}


\begin{document}

\section{The SHiP experiment and its physics case}
SHiP is a newly proposed general purpose fixed target facility at the CERN SPS accelerator. A 400~GeV proton beam will be dumped on a heavy target in order to produce $2\times 10^{20}$ proton-target interactions in 5 years. A dedicated detector downstream of the target will be used to probe a variety of models with light long-lived exotic particles with masses below $\mathcal{O}(10~\mbox{GeV/c}^2)$.
Active neutrino cross-sections and angular distributions will also be studied, thanks to a dedicated detector placed between the target and the hidden sector detector~\cite{Bonivento:2013jag, TP, Alekhin:2015oba}.

SHiP's flagship goal is to use decays of charm and beauty mesons to search for Heavy Neutral Leptons (HNLs), which are right-handed partners of the Standard Model (SM) neutrinos. The existence of such particles is strongly motivated by theory, as
they can simultaneously solve multiple questions unanswered by the SM. In the Neutrino Minimal Standard Model ($\nu$MSM), HNLs allow to explain the baryon asymmetry of the Universe, account for the pattern of
neutrino masses and oscillations and provide a dark matter candidate~\cite{hnl}.


\section{Estimating SHiP's sensitivity}
\subsection{Heavy Neutral Leptons}\label{algorithm}
Using the algorithms described in detail in~\cite{TP}, a Monte Carlo simulation was developed to determine both the rate of HNLs produced at the target and the acceptance of the HNL decay products.  From these estimates, the expected number of events in 5 years of SHiP operation was calculated. 

The number of expected HNLs can be parametrised by factorizing the problem into the HNL production rate and detector acceptance:
\begin{equation}
n(HNL) = N(\mbox{p.o.t}) \times \chi (pp \to HNL) \times \mathcal{P}_{vtx} \times \mathcal{A}_{tot}(HNL\to \mbox{visible})
\end{equation}
where $N(\mbox{p.o.t})=2\times 10^{20}$ is the number of proton-target collisions integrated over five years of operation; $\chi(pp \to HNL)$ is the production rate of HNLs per $pp$ interaction; $\mathcal{P}_{vtx}$ is the probability that an HNL of given mass and couplings decays in SHiP's acceptance; and $\mathcal{A}_{tot}(HNL\to \mbox{visible})$ is the fraction of HNLs decaying in the detector acceptance and resulting in a detectable final state.

The production cross sections of $c$- and $b$- mesons, the main sources of sterile neutrinos at SHiP, are estimated with \textsc{Pythia8}~\cite{Sjostrand:2007gs}, and the HNL production rate takes into account all the dominant, kinematically-allowed decay channels of $D_{(s)}$ and $B_{(s)}$ mesons into sterile neurinos. The widths of these channels are parametrized according to~\cite{Gorbunov:2007ak} as a function of the HNL mass and couplings.
The HNL lifetime is estimated as the sum of the widths of its dominant decay channels ($3\nu$, $\pi^0\nu$, $\pi^\pm\ell$, $\rho^0\nu$, $\rho^\pm\ell$ and $\ell^+ \ell^-\nu$ final states), that are also parametrised according to~\cite{Gorbunov:2007ak}.

\textsc{Pythia8} is used to build a data set containing kinematic information from charmed and beauty mesons produced at the SHiP target. For every meson, all the kinematically allowed decays into massive sterile neutrinos are simulated using the \textsc{ROOT} \textsc{TGenPhaseSpace} class~\cite{Brun199781}. A two-dimensional binned probability density function (PDF) is built out of the resulting HNLs momentum $p$ and polar angle $\theta$, weighted with the branching ratio of the meson decay in which the HNL is produced. This PDF was then used as an estimate of the 4-momentum spectrum of HNLs produced at the SHiP target.
Each bin of this PDF contributes to the total vertex acceptance $\mathcal{P}_{vtx}$ proportionally to its content and to the integral of the decay probability over SHiP's decay volume, as a function of $p$ and $\theta$:
\begin{eqnarray}
\mathcal{P}_{vtx}\left( p,\theta\right) = \int_{\mbox{SHiP}} e^{-l/\gamma c\tau}\, dl\\
\mathcal{P}_{vtx} = \int \mathcal{P}_{vtx}\left( p,\theta  \right)\, dp\, d\theta
\end{eqnarray}

The detectable fraction of HNLs is computed as a combination of the branching ratios and final state acceptances of the HNL decay channels that are kinematically allowed and produce two charged particles in the final state:
\begin{equation}
\mathcal{A}_{tot}(HNL \to \mbox{visible}) = \sum_{i}BR(HNL\to i)\times \mathcal{A}(i)
\end{equation}
where the index $i$ runs over visible final states and each acceptance $\mathcal{A}(i)$ is computed by simulation.

The HNL yield depends on the hierarchy of the active neutrino masses, the HNL couplings to the three SM flavours $U^2_e, U^2_\mu, U^2_\tau$ and the HNL mass. The total coupling $U^2 = \sum_i U^2_i$ and the HNL mass are free parameters in the simulation, while the relative strenghts of the coupling and the hierarchy are fixed and constitute different scenarios which conform to existing theoretical studies~\cite{Gorbunov:2007ak}.
\figurename~\ref{fig:hnl} shows SHiP's sensitivity, along with the foreseen sensitivity of FCC-ee in two realistic configurations (Section~\ref{fcc-ee}), in the parameter space of the $\nu$MSM, for a scenario with inverted hierarchy of active neutrino masses and relative coupling strengths ${U_e^2:U_{\mu}^2:U_{\tau}^2}\sim{48:1:1}$.

\subsection{Dark Photons}
A variety of New Physics models foresee a set of SM-neutral unobserved particles that do not interact with SM particles except through a ``messenger'' particle belonging to the hidden sector. Minimalistic models consist of a $U(1)'$ gauge symmetry in the hidden sector, whose gauge boson $\gamma'$ is called a \textit{dark photon}. If $U(1)'$ is broken by a Higgs-like mechanism, the dark photon can acquire a non-zero mass. Such dark photons may be searched for at SHiP in neutral di-lepton and di-meson final states: $\gamma'$ may mix to the SM photon through loops of particles charged both under $U(1)$ and $U(1)'$. Assuming $\gamma'$ is the lightest particle of the hidden sector, it would then decay to $\ell^+\ell^-$ and $q\bar{q}$ final states through a virtual photon.

A toy simulation was used to estimate SHiP's sensitivity to dark photons. The algorithm is analogous to the one described in Section~\ref{algorithm} and to the method used by the authors of~\cite{darkphoton}.

\begin{figure}[tbp] 
	\centering
	\begin{minipage}{.49\textwidth}
		\includegraphics[width=\textwidth]{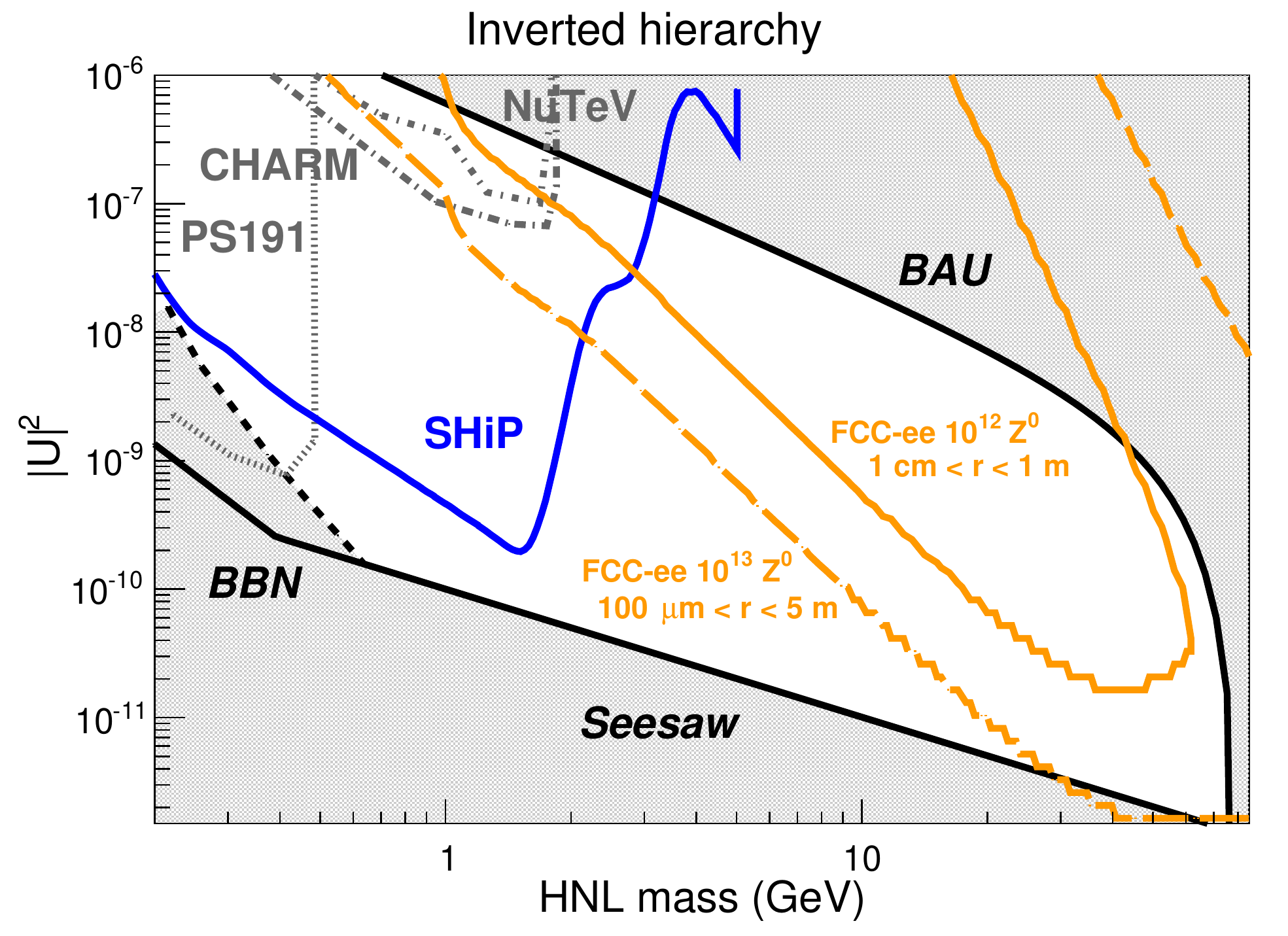}
		\caption{Physics reach in the $\nu$MSM for SHiP and two realistic FCC-ee configurations (see text). Previous searches are shown (dashed lines), as well as the cosmological boundaries of the model (greyed-out areas)~\cite{hnl, Blondel:2014bra}.}\label{fig:hnl}
	\end{minipage}\hfill
	\begin{minipage}{.49\textwidth}
		\includegraphics[width=\textwidth]{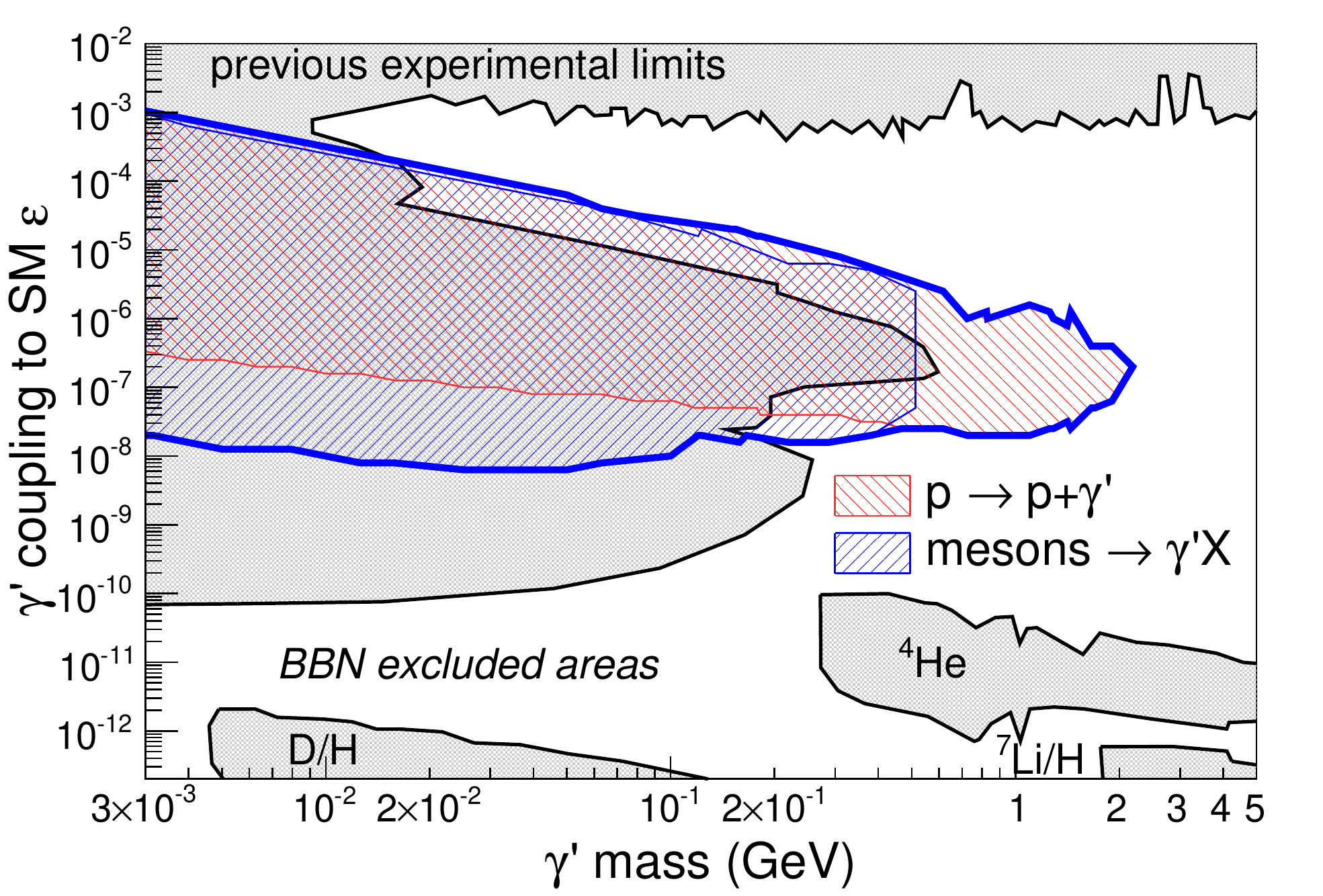}
		\caption{SHiP sensitivity to dark photons produced in proton bremmstrahlung and secondary mesons decays. Previous searches~\cite{Fradette:2014sza} explored the greyed-out area. Low-coupling regions are excluded by Big Bang Nucleosynthesis and SN1987A~\cite{Fradette:2014sza, Kazanas:2014mca}.}\label{fig:dp}
	\end{minipage}
\end{figure}

\figurename~\ref{fig:dp} shows SHiP's sensitivity to dark photons, compared to previous searches. The constraints from supernovae cooling~\cite{Kazanas:2014mca} and big bang nucleosynthesis, and the most up-to-date limits established by previous experiments up to July 2014~\cite{Fradette:2014sza}, are shown in grey.

\section{HNL searches at future colliders}\label{fcc-ee}
A review of possible methods to perform HNL searches at future $e^+e^-$ colliders is given in~\cite{Blondel:2014bra}. This document reports on the sensitivity that can be achieved in direct searches in $Z$ decays.
FCC-ee is a 90-400~GeV high-precision $e^+e^-$ collider aimed at precision tests of the Standard Model, currently studied within the scope of the Future Circular Collider project at CERN. Luminosity studies suggest that the machine could be operated at the $Z$ resonance to produce $10^{12}$ to $10^{13}$ $Z$ bosons per year~\cite{Gomez-Ceballos:2013zzn}.

A method similar to the one outlined in Section~\ref{algorithm} was used to compute the expected number of recorded HNLs. HNL production is assumed to happen in $Z \to \nu\bar{\nu}$ decays with one neutrino kinematically mixing to an HNL. If the accelerator operates at the $Z$ resonance, $Z$ bosons decay at rest and the HNL lifetime is boosted by a factor $ \gamma = {m_Z}/{2 m_N} + {m_N}/{2 m_Z} $.
All $\ell^+\ell^-\nu$ final states are considered detectable. A detector with spherical symmetry is modeled. Backgrounds from $W^*W^*$, $Z^*Z^*$ and $Z^*\gamma^*$ processes can be suppressed by requiring a displaced secondary vertex.

The minimum and maximum displacements of the secondary vertex in FCC-ee, referred to as $r$ in \figurename~\ref{fig:hnl}, depend on the resolution and dimensions of the tracking system. The two FCC-ee sensitivity contours shown in \figurename~\ref{fig:hnl} represent two realistic scenarios~\cite{Blondel:2014bra}.
The production of $10^{12}$ ($10^{13}$) $Z$ bosons is assumed.
$Z$ bosons are allowed to decay in $\nu+ HNL$ pairs, followed by the subsequent $HNL \to \ell\ell\nu$ decay. Secondary vertices between 1~cm (100~$\mu$m) and 1~m (5~m) from the interaction point are selected.

This work shows that the SHiP experiment can improve the current limits on Heavy Neutral Leptons by several orders of magnitude, scanning a large part of the parameter space below the $B$ meson mass. Similarly, SHiP can greatly improve present constraints on dark photons. Right-handed neutrinos with larger mass can be searched for at a future $Z$ factory. The synergy between SHiP and a future $Z$ factory would allow the exploration of most of the $\nu$MSM parameter space for sterile neutrinos.



\acknowledgments

This work would not have been possible without the precious theory support by M. Shaposhnikov. We thank A. Blondel for useful discussions about the FCC-ee project. We are indebted to all our colleagues from the SHiP collaboration for invaluable comments and enthusiasm.

\bibliographystyle{JHEP}
\bibliography{bib}

\end{document}